\documentclass[useAMS,usenatbib]{mn2e}

\usepackage{aas_macros}
\usepackage[footnotesize]{subfigure}
\usepackage{graphicx}

\newcommand\gtsim{\mathrel{\lower0.6ex\hbox{$\buildrel {\textstyle >}\over {\scriptstyle \sim}$}}}
\newcommand\ltsim{\mathrel{\lower0.6ex\hbox{$\buildrel {\textstyle <}\over {\scriptstyle \sim}$}}}
\newcommand\qso{SDSS~J0924+0219 }
\newcommand\mg{MG~0414+0534 }
\newcommand\ec{QSO~2237+0305}
\newcommand\he{HE~1104-1805}

\title[The accretion disc in the quasar SDSS~J0924+0219]{The accretion disc in the quasar SDSS~J0924+0219\thanks{This paper includes data gathered with the 6.5 meter Magellan Telescopes located at Las Campanas Observatory, Chile.}}

\author[D. J. E. Floyd, N. F. Bate and R. L. Webster]
{D. J. E. Floyd$^{1,2}$\thanks{E-mail: dfloyd@lco.cl (DJEF), 
nbate@physics.unimelb.edu.au (NFB), rwebster@unimelb.edu.au (RLW)}
N. F. Bate$^{2}$\footnotemark[2] and R. L. Webster$^{2}$\footnotemark[2]\\
$^{1}$OCIW, Las Campanas Observatory, Casilla 601, Colina El  Pino, La Serena, Chile\\
$^{2}$School of Physics, The University of Melbourne, Parkville, Vic, 3010, Australia}
\begin{document}

\date{Accepted .... Received 02/04/2009; in original form 02/04/2009}

\pagerange{\pageref{firstpage}--\pageref{lastpage}} \pubyear{2009}

\maketitle

\label{firstpage}

\begin{abstract}
We present single-epoch multi-wavelength optical-NIR observations of the ``anomalous'' lensed quasar SDSS~J0924+0219, made using the Magellan 6.5-metre Baade telescope at Las Campanas Observatory, Chile. The data clearly resolve the anomalous bright image pair in the lensed system, and exhibit a strong decrease in the anomalous flux ratio with decreasing wavelength. This is interpreted as a result of microlensing of a source of decreasing size in the core of the lensed quasar. We model the radius of the continuum emission region, $\sigma$, as a power-law in wavelength, $\sigma\propto\lambda^\zeta$. We place an upper limit on the Gaussian radius of the $u^\prime$-band emission region of $3.04\times10^{16} h_{70}^{-1/2} (\langle M \rangle/M_{\odot})^{1/2}$~cm, and constrain the size-wavelength power-law index to $\zeta<1.34$ at 95\% confidence. These observations rule out an alpha-disc prescription for the accretion disc in SDSS~J0924+0219 with 94\% confidence.
\end{abstract}


\section{Introduction}
\label{sec-intro}
Half a century on from the discovery of quasars our paradigm for understanding their central engine remains poorly physically motivated and only very weakly observationally constrained~\citep{blaes07}. While the standard accretion model has met significant success in describing Cataclysmic Variables and Young Stellar Objects, it does a poor job when dealing with accretion onto black holes, and is thermally and viscously unstable in situations where radiation pressure dominates gas pressure (see the review by~\citealt{blaes02}). 
AGN accretion discs are impossible to observe directly, having typical angular sizes on the order of micro-to-nano arcseconds. 

Microlensing is an established technique for obtaining high spatial resolution measurements of lensed objects (e.g.~\citealt{WP91,RB91,WMS95,WWTM00,SW02} ), but until recently the technique was not refined enough to place meaningful statistical constraints on lensed quasar emission regions. \citet{bate+07} presented a technique for calculating source size probabilities using Monte Carlo simulations of magnification maps at the minimum and saddle-point image of an ``anomalous'' quadruply-imaged lensed quasar, which allows us to place far tighter constraints on the central source sizes of quasars.

In an AGN accretion disc, gravitationally bound material spirals inwards onto a central black hole. Gravitational compression heats the material, which then radiates. In order to fall inwards, matter must lose not only energy but also angular momentum, while the total angular momentum of the system is conserved. The central problem in accretion discs is one of angular momentum transport outwards in the disc, in order that material can continue to plunge inwards. See the excellent reviews by~\citet{blaes02} and~\citet{balbus03} for detailed information.
While molecular viscosity helps heat up the disc and thus radiate energy, it is insufficient to transport the angular momentum outwards. The idea of an anomalous stress introduced by nonlinear physics (``turbulence'') dates back to~\citet{SS73}. A longstanding problem has been how to introduce such a turbulence to what is essentially a stable laminar flow -- the Rayleigh Stability criterion excludes any hydrodynamic mechanism (e.g.~\citealt{FKR02,balbus03,blaes02}). A range of theoretical models based on different physical mechanisms have predicted different radial emission profiles.

\cite{SS73} first invoked a non-linear turbulence to produce an additional stress in the disc, parameterised by a coefficient $0 < \alpha < 1$. 
Based solely on energetics and conservation of mass (ignoring the issue of angular momentum transport entirely), one can derive a simple radial temperature profile, $T_c \propto r^{-p}$, with $p=3/4$ (see e.g.~\citealt{PR72,SS73,gaskell08}). If the emergent spectrum is the super-position of blackbodies generated locally, $T \propto \lambda^{-1}_{max}$ (e.g.~\citealt{FKR02}). This gives a spectral profile

\begin{equation}
r \propto \lambda^{1/p} \propto \lambda^{\zeta}
\label{ssdisc}
\end{equation}
where $\zeta = 4/3$. This is what you get from a simple \citet{SS73} treatment of the disc. \citet{gaskell08} suggests the opposite extreme, in which all energy simply comes from the outward flow from hotter to cooler regions. With a single central source, and thus a flux that falls off as $r^{-2}$, the effective temperature drops off as $T\propto r^{-1/2}$, giving $p=0.5$, $F_\nu\propto\nu^{-1}$ and $\zeta=2$. The ``slim disc''~\citep{abramowicz+88} is an example of a model that gives rise to such a temperature profile. \cite{gaskell08} also calculates the value of $p$ required in order to fit a generic observed quasar spectral index in the optical--UV: $F_\nu\propto\nu^{-0.5}$. This requires a value of $p=0.57$, and thus $\zeta=1.75$.
It has also been suggested that magneto-rotational instability (MRI) might be a generic source of turbulence in discs~\citep{BH91}.  \citet{agolkrolik00} present an example of such a model that assumes a strong magnetic connection between black hole and disc that essentially ``spins up'' the disc from the black hole's own rotation. This gives rise to a temperature profile with $p=7/8$, or a wavelength profile index, $\zeta = 8/7$.


Anomalous lensed quasars have an image pair in which one of the images is unusually (``anomalously'') dim, with the evidence strongly suggesting that the anomaly is caused by microlensing. Observations have shown that the brightness anomaly increases with decreasing wavelength (e.g.~\citealt{WP91,inada+03,bate+08}).  Some studies of such sources have attributed the anomaly to dust, e.g.~\citet{lawrence+95a,mcleod+98,OGM08}. Most of these studies were carried out before the relevant lensing galaxy redshifts were known, and more recent observations generally cast doubt on the dust hypothesis. In most (eight out of ten) known cases~\citep{pooley+07} the anomalously dim image lies at the saddle-point of the Fermat light-travel time surface as predicted by lensing theory. It is difficult to achieve this with dust ($P_{\geq8/10}=0.04$ from the binomial distribution). 
Furthermore, X-ray observations of the quadruply lensed quasar 1RXS J1131--1231 show a very low absorbing column density, ruling out differential absorption in at least this case~\citep{blackburne+06}. Millilensing by galactic substructure has also been explored, but would not produce the observed polychromatic effect~\citep{MS98, MM01, C02, DK02}. Of course gravitational lensing is a fundamentally achromatic process at any scale, and the microlensing interpretation introduces a chromatic effect by virtue of the lensed source size varying with wavelength. A strongly anomalous radio flux ratio would provide strong evidence for millilensing, but none has been observed to date.

Assuming that the anomaly in our first observed object (MG~0414+0534) is entirely due to microlensing,~\citet{bate+08} placed some of the strongest constraints yet on accretion disc size, but placed only weak constraints on the emission mechanism, excluding the simple turbulence-driven~\citet{SS73} model at only the 68\% level. Here we present our second case, SDSS~J0924+0219. We are able to place much stronger constraints on the observed index and effectively rule out the~\citet{SS73} disc scenario in this source. Note that similar work can be done using long-term monitoring of non-anomalous lensed quasars~\citep{eigenbrod+08b}. The advantage of using anomalous lensed quasars is that the close proximity of the anomalous image pair guarantees a short relative time delay between the two images. Thus observing the sources close together in time means we are not seeing the source at two different epochs. Our technique is applicable to only a small subset of known lenses, but for this subset, using single epoch data, tight constraints can be obtained, that in other sources would take decades.


We assume throughout this paper a cosmology with $H_0=70$~km~s$^{-1}$~Mpc$^{-1}$, $\Omega_m = 0.3$ and $\Omega_\Lambda= 0.7$.

\section{Observations}
SDSS~J0924+0219  was the second lensed quasar to be discovered by the Sloan Digital Sky Survey (SDSS). It is the most anomalous lensed quasar known. It  was previously imaged on Magellan (MagIC) by~\cite{inada+03} in 0\farcs55-0\farcs75 seeing as a follow-up to initial discovery by SDSS. The anomalous pair of images, labelled A (brightest image) and D, are weakly resolved in these data. We obtained the original MagIC imaging data taken by~\citet{inada+03} in order to work directly on the raw images. 
There are no existing radio, mm nor sub-mm detections of the source, and no mentions of dust in this source in the literature.
\cite{keeton+06} presented HST/ACS and NICMOS imaging and spectroscopy of the source, showing that the anomaly is present in both the continuum and the broad emission line flux ratios. Chandra soft X-ray measurements also show a strong anomaly~\citep{pooley+07}. The redshift of the quasar is $z_s=1.524$~\citep{inada+03} and  the redshift of the lens galaxy is $z_l=0.394$~\citep{eigenbrod+06a}.

\begin{figure*}
\includegraphics[width=160mm]{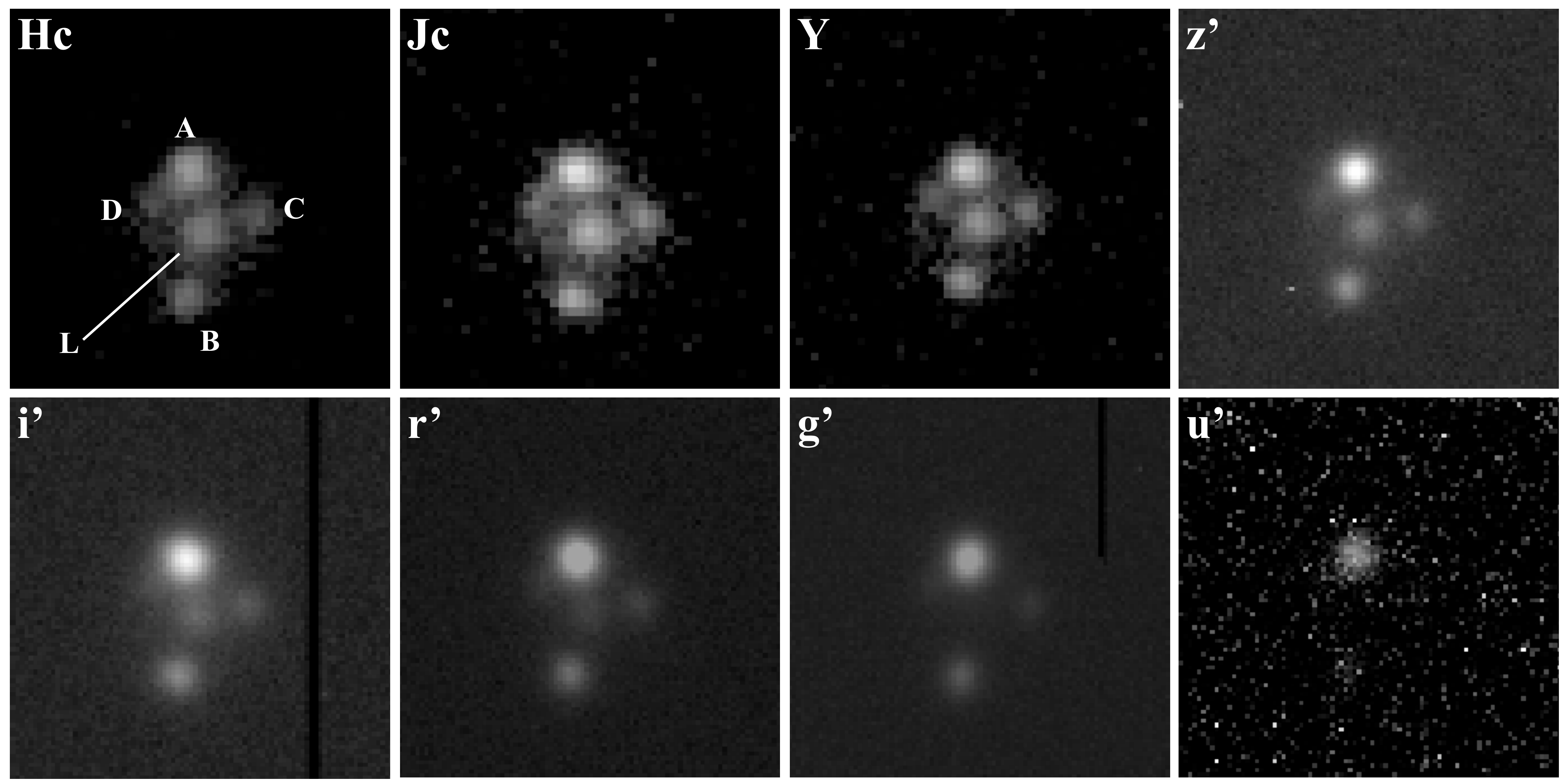}
\caption{\label{fig-images} Magellan IMACS and PANIC imaging of SDSS~J0924+0219 taken on 2008 March 21. 
The anomaly between $A$ and $D$ becomes more significant as we observe blueward. The $u^\prime$ image was not used and is included for completeness only.}
\end{figure*}

In this paper we present a third epoch of optical-NIR observations, taken with the Magellan 6.5-metre Baade telescope on 2008 March 21 in $0\farcs3 - 0\farcs45$~seeing. These observations complement the earlier two epochs of optical-IR data, spanning a larger wavelength range (0.4 -- 1.6$\mu$m) near-simultaneously, and in greater spectral resolution (seven bands) than any earlier dataset. Observations were made using IMACS (in f/2 mode) in SDSS $g^\prime, r^\prime, i^\prime$ and $z^\prime$ bands, and PANIC in $Y$, $J$ and $H$. Integration times were approximately five minutes per filter. Images are provided in Figure~\ref{fig-images}. Photometric calibrations were performed using three standard star observations over the course of the night in the optical, and two in the NIR. We note that any biases in the calibration will not affect the $D/A$ ratios, and thus our modelling of the data, which depends only on the relative fluxes of the anomalous image pair $A$ and $D$. We fit Gaussians to each quasar image, beginning with an assumption of a round, Gaussian PSF, and then allow the ellipticity to vary as a free parameter. We initially used the HST data to constrain the position of each component in each of our observations. In the final fit we freed up position as well, resulting in a 5-D fit (x, y, amplitude, FWHM, and ellipticity) for each component using a downhill gradient technique. The apparent magnitudes of the four lensed images and the lensing galaxy are provided in Table~\ref{tab-mag}. Flux ratios in each filter between the two images of interest, $F_{obs} = {D}/{A}$, are provided in Table~\ref{tab-ratio}. This table contains flux ratios from the observations presented in this paper, as well as Chandra data~\citep{pooley+07}, HST observations obtained from the CASTLES Survey web page\footnote{http://cfa-www.harvard.edu/castles/} and the earlier Magellan observations of~\citet{inada+03}.

\begin{table*}
\centering
  \caption{\label{tab-mag} New SDSS~J0924+0219 photometry from Magellan IMACS/PANIC: Apparent magnitudes of the lensed images (A, D, B, and C) and the lensing galaxy (G) in SDSS~J0924+0219. Observations were taken in seven filters with the IMACS and PANIC instruments on the Magellan 6.5-m Baade telescope, 2008 March 21. Typical photometric errors are 3.5\% in the NIR and 6\% in the optical.}
  \begin{tabular}{lrrrrr}
  \hline
  \hline
  Filter & A & D & B & C & G \\
  \hline
  $H$        & $17.39\pm0.04$ & $18.99\pm0.04$ & $18.34\pm0.04$ & $18.80\pm0.04$& $17.39\pm0.04$\\
  $J$        & $18.24\pm0.04$ & $20.29\pm0.04$ & $19.44\pm0.04$ & $20.02\pm0.04$& $18.16\pm0.04$\\
  $Y$        & $18.30\pm0.04$ & $20.46\pm0.04$ & $19.51\pm0.04$ & $20.09\pm0.04$& $19.07\pm0.04$\\
  $z^\prime$  & $19.84\pm0.06$ & $21.65\pm0.06$ & $20.98\pm0.06$ & $21.65\pm0.06$& $21.15\pm0.06$\\
  $i^\prime$  & $19.54\pm0.06$ & $21.54\pm0.06$ & $20.77\pm0.06$ & $21.40\pm0.06$& $21.06\pm0.06$\\
  $r^\prime$  & $19.85\pm0.06$ & $22.32\pm0.06$ & $21.31\pm0.06$ & $22.59\pm0.06$& $20.41\pm0.06$\\
  $g^\prime$  & $20.14\pm0.06$ & $22.90\pm0.06$ & $21.57\pm0.06$ & $22.72\pm0.06$& $21.63\pm0.06$\\
  \hline
\end{tabular}
\end{table*}

\section{Modelling}
We parameterise the radius of the emission region $\sigma$ as a power-law in wavelength in order to test against various accretion-disc models:

\begin{equation}
\sigma = \sigma_0 \left(\frac{\lambda}{\lambda_0}\right)^\zeta
\label{powerlaw}
\end{equation}
We use the modelling technique described by~\citet{bate+07}, and applied to MG0414+0534~\citep{bate+08}. To re-iterate briefly, we begin with a macro-lensing model obtained from~\citet{keeton+06}. This provides the convergence $\kappa$ and shear $\lambda$ at each image position. We consider two distinct components to convergence -- a continuous component $\kappa_c$ and clumpy, compact stellar component $\kappa_\star$. Caustic networks are then constructed using an inverse ray-shooting method~\citep{wambsganss+90a}. The microlens population is drawn from a Salpeter mass function, $dN/dM \propto M^{-2.35}$, with $M_{max}/M_{min} = 50$. Physical sizes are scaled by the average microlens Einstein radius projected onto the source plane, $\eta_{0} = 5.78\times10^{16} h_{70}^{-1/2}\left(\langle M \rangle / M_{\odot}\right)^{1/2}$~cm. We also performed simulations using a fixed mass microlens population (at the average mass of the Salpeter function above).

Gaussian source surface brightness profiles \textit{b} of various characteristic radii $\sigma$ are convolved with the caustic networks in the image plane:

\begin{equation}
b = \rmn{exp}\left(-\frac{x^2+y^2}{2\sigma^2}\right)
\label{gaussian}
\end{equation}
We then construct a library of simulated observations of flux ratio as a function of source radius. These are converted to flux ratio as a function of wavelength using Equation \ref{powerlaw}. Comparing this library of simulations with the observational data allows us to build up three dimensional probability distributions for smooth matter fraction, radius of the inner-most emission region $\sigma_0$, and power-law index $\zeta$. 
Smooth matter was allowed to vary between 0\% and 99\% in 10\% increments. We obtain the results presented in this paper by marginalising over smooth matter fraction:

\begin{equation}
\frac{d^2P}{d\zeta d\sigma_0} = \int\frac{d^3P}{ds d\zeta d\sigma_0}ds
\end{equation}

Source sizes were 0.05 $\eta_0$ to 2.00 $\eta_0$, in steps of 0.05 $\eta_0$. The upper end is large enough that microlensing should no longer be significant. The lower end is just above the resolution limit of our simulations, and is equivalent to the Schwarschild radius $R_s=2GM/c^2$ of a $10^9~M_\odot$ black hole. The mass of the black hole in \qso has been estimated at between $1.1\times10^8 M_\odot$~\citep{peng+06} and $2.8\times10^8 M_\odot$~\citep{morgan+06}, so the lower limit of our calculation is close to the expected innermost stable circular orbit of a nonrotating black hole, $3R_s=6GM/c^2$.

As discussed by~\cite{keeton+06}, microlensing alone is sufficient to explain the observed anomaly.
However,~\citet{eigenbrod+06a} have argued for the presence of CDM substructure in the vicinity of image $D$. Thus millilensing may also be present. However, this would not produced the observed chromatic effect, rather it would offset the total observed anomaly. 

\section{Results}
\begin{figure}
\includegraphics[width=80mm]{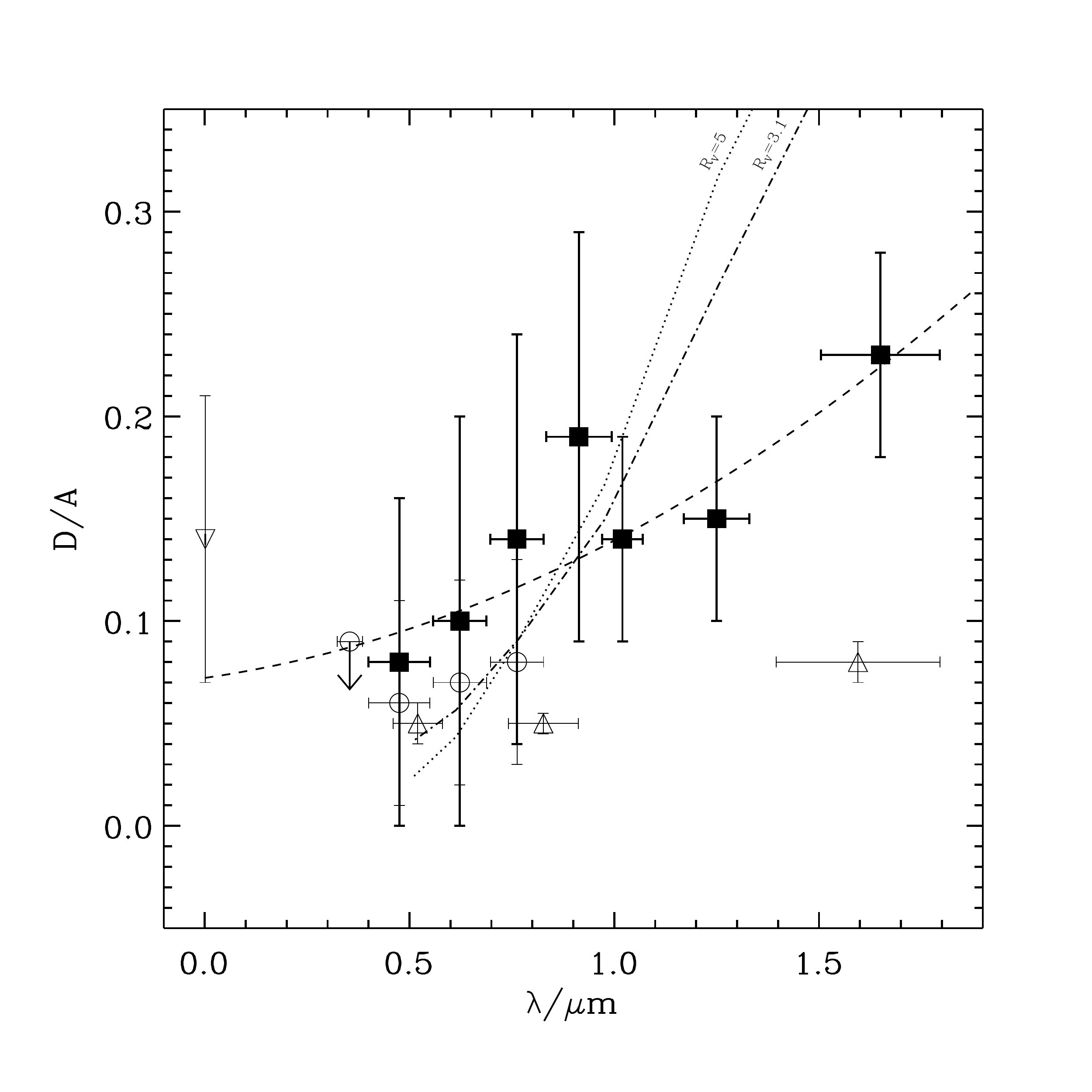}
\caption{\label{fig-dust} Anomalous flux ratio with wavelength for 
this study (filled squares) and for earlier data: 
Circles~\citet{inada+03}; Triangles~\citet{keeton+06}; Down-pointing 
triangle~\citep{pooley+07}. The dashed line shows the best quadratic 
fit to our new data only. For reference, we model the ratios 
if we assume that $D$ is simply extinguished by dust at the redshift of the lens, $z_l=0.394$~\citep{inada+03}, with no microlensing. Two dust models are shown~\citep{mathis90} with $R_V=3.1$ 
(dotted line -- typical diffuse ISM) and $R_V=5$ (dot-dashed line -- 
dense dust clouds). Partial extinction would result in the 
lines moving to the right, with the same slopes.
Millilensing can help explain the strongly anomalous NIR flux ratio, but cannot explain the slope.}
\end{figure}

\begin{figure}
\includegraphics[width=80mm,angle=0]{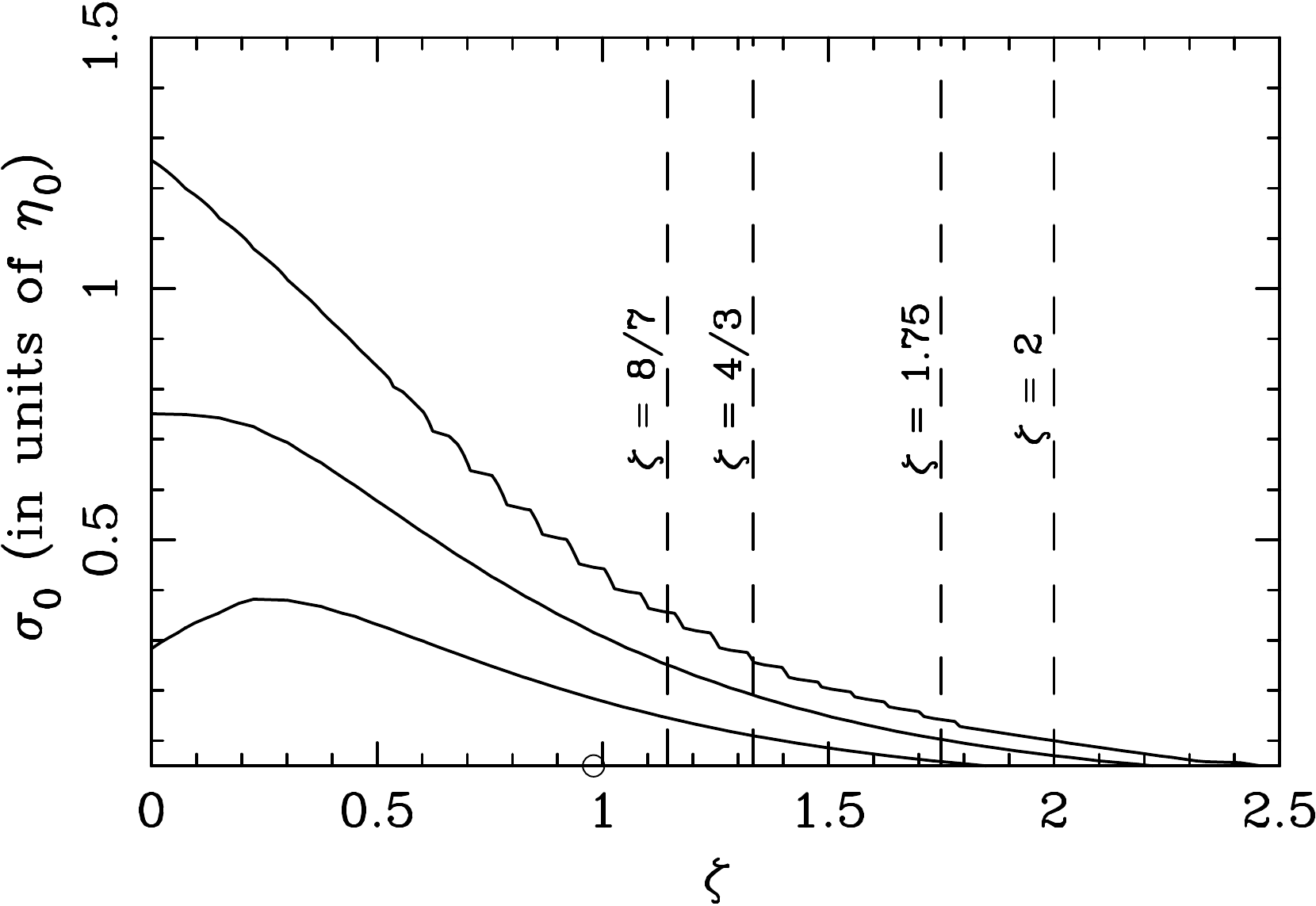}
\includegraphics[width=80mm,angle=0]{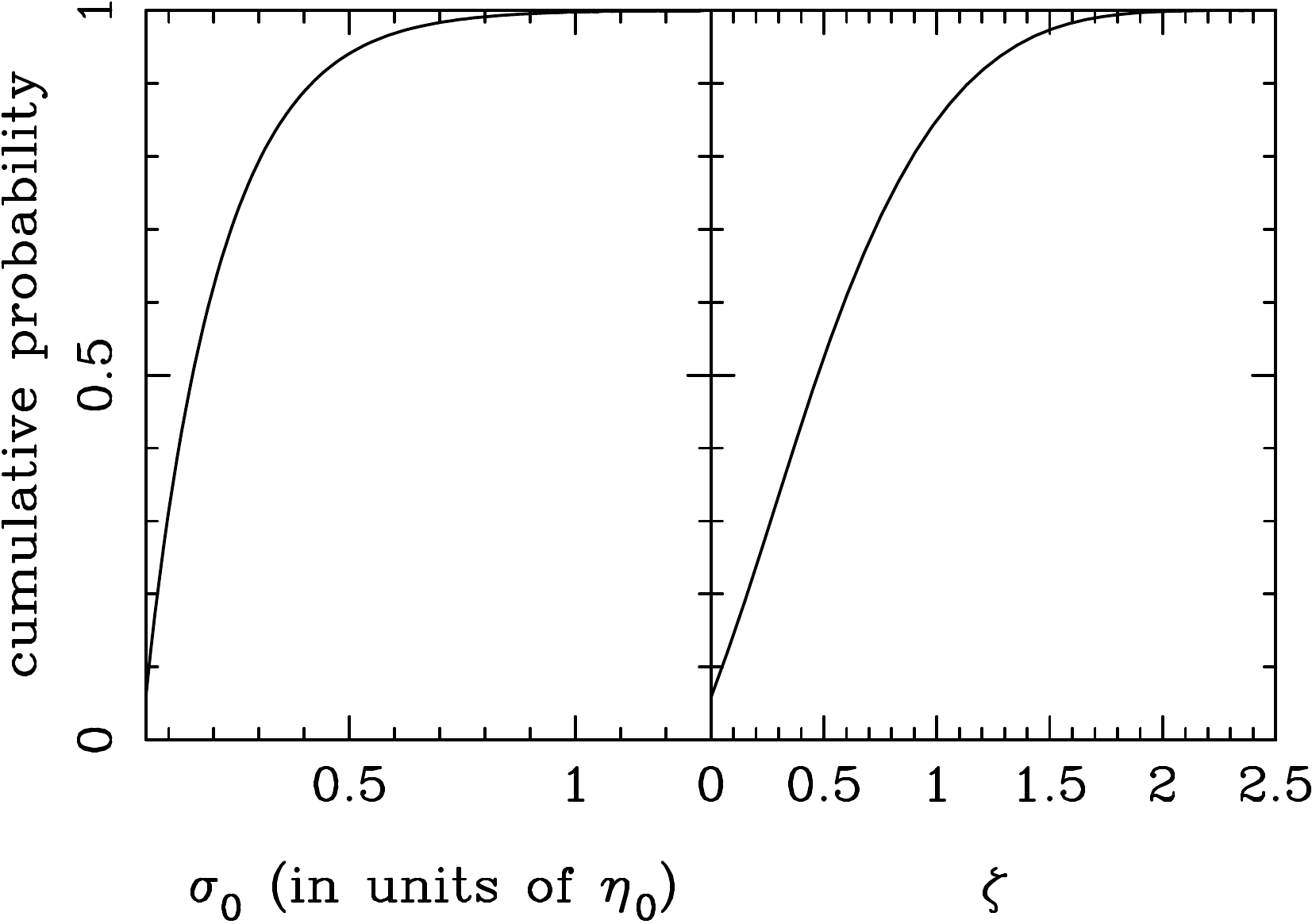}
\caption{\label{fig-res} Top: Probability distributions for the radius of 
the $u^\prime$-band emission region, $\sigma_0$, in units of the average 
Einstein Radius $\eta_0$, and the power-law index $\zeta$, based
 on our new Magellan data only. Contours are shown at 1, 2, and 3$\sigma$.
The dashed lines illustrate four accretion disc models, as described in the main text. 
The small circle indicates the location of the maximum in the probability distribution.
Bottom: Cumulative probability distributions for $\sigma_0$ (left) and $\zeta$ (right).}
\end{figure}

Our newly acquired images of SDSS~J0924+0219  are presented in Figure~\ref{fig-images}. In Table~\ref{tab-mag} we present our PSF photometry. Table~\ref{tab-ratio} shows that the flux ratio between images $D$ and $A$ grows larger as wavelength increases. We also show all published values of the anomalous flux ratio in the literature, from X-ray to NIR. The anomalous flux ratios are plotted against wavelength in Figure~\ref{fig-dust}. We assume that this anomaly is solely due to microlensing, and apply microlens Monte Carlo simulations as described above to  return a joint probability distribution in source size, radial wavelength power index and smooth matter fraction in the lens. Marginalising over the latter we obtain the probability distribution presented in Figure~\ref{fig-res}. Cumulative probability distributions in source size $\sigma$ and radial wavelength index $\zeta$ are also shown. We model the size of the continuum emission region, $\sigma$, as a power-law in wavelength, $\sigma\propto\lambda^\zeta$. We place an upper limit on the Gaussian radius of the emitting region of $3.04\times10^{16} h_{70}^{-1/2} (\langle M \rangle/M_{\odot})^{1/2}$~cm ($11.73 h_{70}^{-1/2} (\langle M \rangle/M_{\odot})^{1/2}$ light days) in $u^\prime$-band, and constrain the power-law index to $\zeta<1.34$ at 95\% confidence.

\begin{figure}
\includegraphics[width=80mm,angle=0]{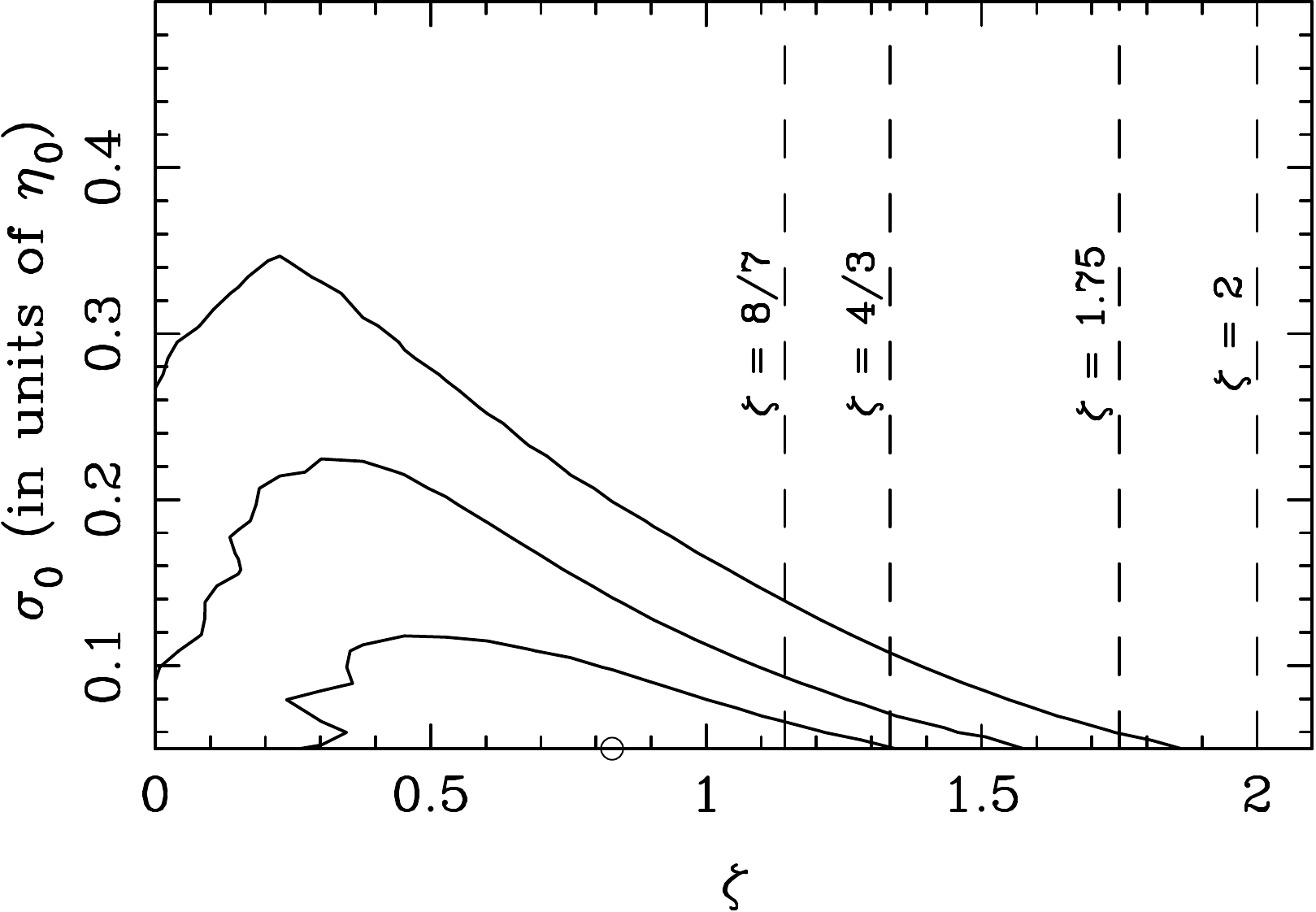}
\includegraphics[width=80mm,angle=0]{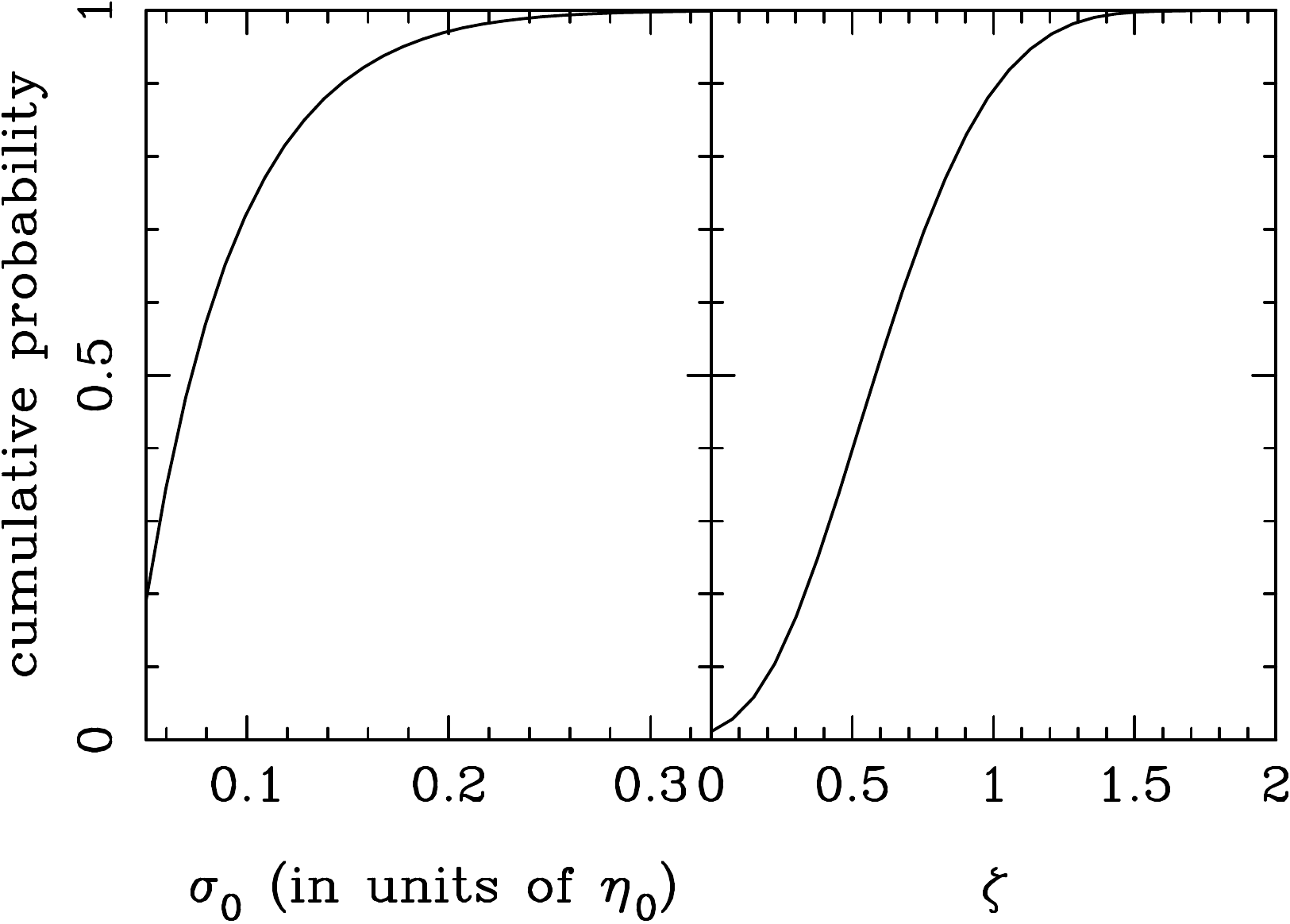}
\caption{\label{fig-resall} As for Figure~\ref{fig-res} but for all available data (this paper, Inada et al.\ 2003, Keeton et al.\ 2006). The top panel shows the probability distributions for $u^\prime$ source radius, $\sigma_0$ and the power-law index. The small circle indicates the location of the maximum in the probability distribution.
The bottom panel shows the cumulative probability distributions for $\sigma_0$ (left) and $\zeta$ (right). The alpha-disc prescriptions ($\zeta>4/3$) are excluded at the $97\%$ level by this combined dataset.}
\end{figure}

If we add in all the observational data (assuming that each epoch of data applies an independent constraint to the accretion disc structure) we obtain the slightly tighter probability distribution shown in Figure~\ref{fig-resall}. Together they constrain the radius of the emitting region to be $ \ltsim 1.02\times10^{16} h_{70}^{-1/2} (\langle M \rangle/M_{\odot})^{1/2}$~cm ($3.95 h_{70}^{-1/2} (\langle M \rangle/M_{\odot})^{1/2}$ light days) in $u^\prime$-band, and the power-law index to $0.06<\zeta<1.24$ at 95\% confidence.

We have presented results obtained using a Salpeter microlens mass function as we feel they are most generally applicable. Switching to a fixed mass microlens population (mean mass of the Salpeter function) results in near-identical constraints on $\zeta$ ($\zeta < 1.32$), but with slightly tighter constraints on the actual source size: $0.31\eta_0$, or $1.77\times10^{16}h_{70}^{-1/2} (\langle M \rangle/M_{\odot})^{1/2}$~cm in $u^\prime$-band , or $6.85 h_{70}^{-1/2} (\langle M \rangle/M_{\odot})^{1/2}$ light days at 95\%, using the Magellan data only. 

\begin{table}
\centering
\caption{\label{tab-ratio} Central wavelengths and observed (anomalous) flux ratios $F_{obs}$ between images $A$ and $D$ in each filter. The 2008 March 21 observations were taken with the IMACS and PANIC instruments on the Magellan 6.5-m Baade telescope. $^{(a)}$2001 December 15 data were taken using MagIC on Baade~\citep{inada+03}. $^{(b)}$2003 data were taken using HST/NICMOS and WFPC2 for the CASTLES survey~\citep{keeton+06}. $^{(c)}$2005 data were taken using Chandra~\citep{pooley+07}.} 
\begin{tabular}{lrll}
\hline
\hline
Filter & $\lambda_c$ (\AA) & $F_{obs}={D}/{A}$ & Date\\
\hline
$H$			& $16500\pm1450$	& $0.23\pm0.05$	& 2008 March 21 \\
$J$			& $12500\pm800$	& $0.15\pm0.05$	& 2008 March 21 \\
$Y$			& $10200\pm500$	& $0.14\pm0.05$	& 2008 March 21 \\
$z^\prime$	& $9134\pm800$	& $0.19\pm0.10$	& 2008 March 21 \\
$i^\prime$	& $7625\pm650$	& $0.16\pm0.10$	& 2008 March 21 \\
$r^\prime$	& $6231\pm650$	& $0.10\pm0.10$	& 2008 March 21 \\
$g^\prime$	& $4750\pm750$	& $0.08\pm0.08$	& 2008 March 21 \\
$i^{\prime(a)}$	& $7625\pm650$	& $0.08\pm0.05$ 	& 2001 December 15 \\
$r^{\prime(a)}$	& $6231\pm650$	& $0.07\pm0.05$	& 2001 December 15 \\
$g^{\prime(a)}$	& $4750\pm750$	& $0.06\pm0.05$	& 2001 December 15 \\
$u^{\prime(a)}$	& $3540\pm310$	& $<$0.09			& 2001 December 15 \\
F160W$^{(b)}$	& $15950\pm2000$	& $0.08\pm0.01$	& 2003 November 18 \\
F814W$^{(b)}$	& $8269\pm850$	& $0.05\pm0.005$	& 2003 November 18--19 \\
F555W$^{(b)}$	& $5202\pm600$	& $0.05\pm0.01$	& 2003 November 23 \\
0.5-8keV$^{(c)}$ & $12\pm11$	& $0.14^{+0.07}_{-0.06}$ 	& 2005 Feb 24 \\
\hline
\end{tabular}
\end{table}

\section{Discussion}
\citet{keeton+06} established that microlensing is present in this system, based on the difference in the continuum and broad emission line flux ratios and its photometric variability on the timescale of a decade~\citep{kochanek05}. \citet{eigenbrod+06a} see rapid (15 days) and asymmetric changes in the emission lines of the quasar images, and no strong changes in the continuum. 
We find the time delay between the two close images $A$ and $D$ to be less than a day, using the adopted macrolens model of~\citet{keeton+06} and the GRAVLENS software~\citep{gravlens}. See also table 3 of~\citet{eigenbrod+06a} who present the time delays for a variety of macrolensing models, all of which are consistent with $t_AD<1$~day. Our results therefore simply rely on the choice of the model parameters for the particular images in this source. 

Unfortunately in their follow-up Keck spectra~\citet{inada+03} observed $A$ and one of the other (non-anomalous) images, $B$. 
However,~\citet{keeton+06} present spatially resolved HST spectroscopy of the entire system. The authors note that there are differences between the emission-line and continuum flux $D/A$ ratios. The $D/A$ ratio is 0.1 in the broad Ly$\alpha$ line but 0.05 in the associated continuum. They argue that microlensing can account for both the continuum and emission-line flux ratios, if the broad emission line region is comparable in size to the Einstein radii of the microlenses. Comparing our flux ratios to those of Inada et al.\ and Keeton et al.\ confirms that we are indeed seeing the effects of microlensing, since the values have changed significantly on a 5 year time scale. However, all the existing NIR ratios also show a significant anomaly. There are no radio detections for this source and  without high resolution radio imaging we cannot rule out millilensing entirely. A strong anomalous ratio $D/A$ in the radio would strongly confirm the presence of millilensing. If confirmed this would shed interesting light on the nature of dark matter in the lens, while not strongly affecting our conclusions, since the observed chromaticity cannot be explained by millilensing. 

What happens if we assume that the anomaly is partially or wholly dust? There are three strong arguments against differential extinction by dust in this source. Firstly, we would expect a far smaller effect in the NIR, where little extinction would be expected. To get such strong NIR extinction, we would expect extraordinarily large extinction in the optical. This is illustrated in Figure~\ref{fig-dust}. No physical dust model exists that can produce the observed effect. Dust cannot explain the shape of the observed drop off with wavelength. Secondly, in eight out of ten known anomalous lenses, it is the saddle image which is fainter (see~\citealt{pooley+07}).  The probability of dust extinction would be the same for each image.  And finally, in one other case, extinction by dust has been has been shown to be negligible from the x-ray observations~\citep{blackburne+06}.
If we assume that millilensing accounts for all of the observed effect in the NIR, and that dust caused the remaining differential extinction, we would still see a far steeper drop off in the flux of image $D$ to the blue. 


\section{Conclusions}
We have modelled the size of the continuum emission region, $\sigma$, as a power-law in wavelength, $\sigma\propto\lambda^\zeta$ for the lensed quasar \qso. Using our new Magellan data alone we place an upper limit on the Gaussian radius of the $u^\prime$-band emission region of $3.04\times10^{16} h_{70}^{-1/2} (\langle M \rangle/M_{\odot})^{1/2}$~cm (a little under $12 h_{70}^{-1/2} (\langle M \rangle/M_{\odot})^{1/2}$ light days), and constrain the power-law index to $\zeta<1.34$ at 95\% confidence. This is tighter than  the constraint placed on MG0414+0534~\citep{bate+08}, and excludes the accretion model of~\cite{gaskell08} at greater than the 99\% confidence level. The simple~\cite{SS73} disc is excluded at lower confidence (94\%). Combining our data with earlier epoch HST (ACS and NICMOS) and Magellan (MagIC) data allow us to exclude all alpha-disc models~\citep{SS73, gaskell08} at 97\% confidence, and constrains the source size to $\ltsim 4 h_{70}^{-1/2} (\langle M \rangle/M_{\odot})^{1/2}$ light days.

We explore extinction of image $D$ by dust, and find a number of arguments against this possibility. This hypothesis could be comprehensively tested with high resolution NIR spectroscopic observations aimed at resolving any metal absorption lines. 
Millilensing remains a possibility that would explain the extremely anomalous NIR flux ratio, and which could be ruled out  with VLBI radio observations. However millilensing cannot explain the observed chromaticity of the anomalous flux ratio, and even millilensing and  dust together do not explain the observed spectral shape of the source.

\begin{figure}
\includegraphics[width=85mm]{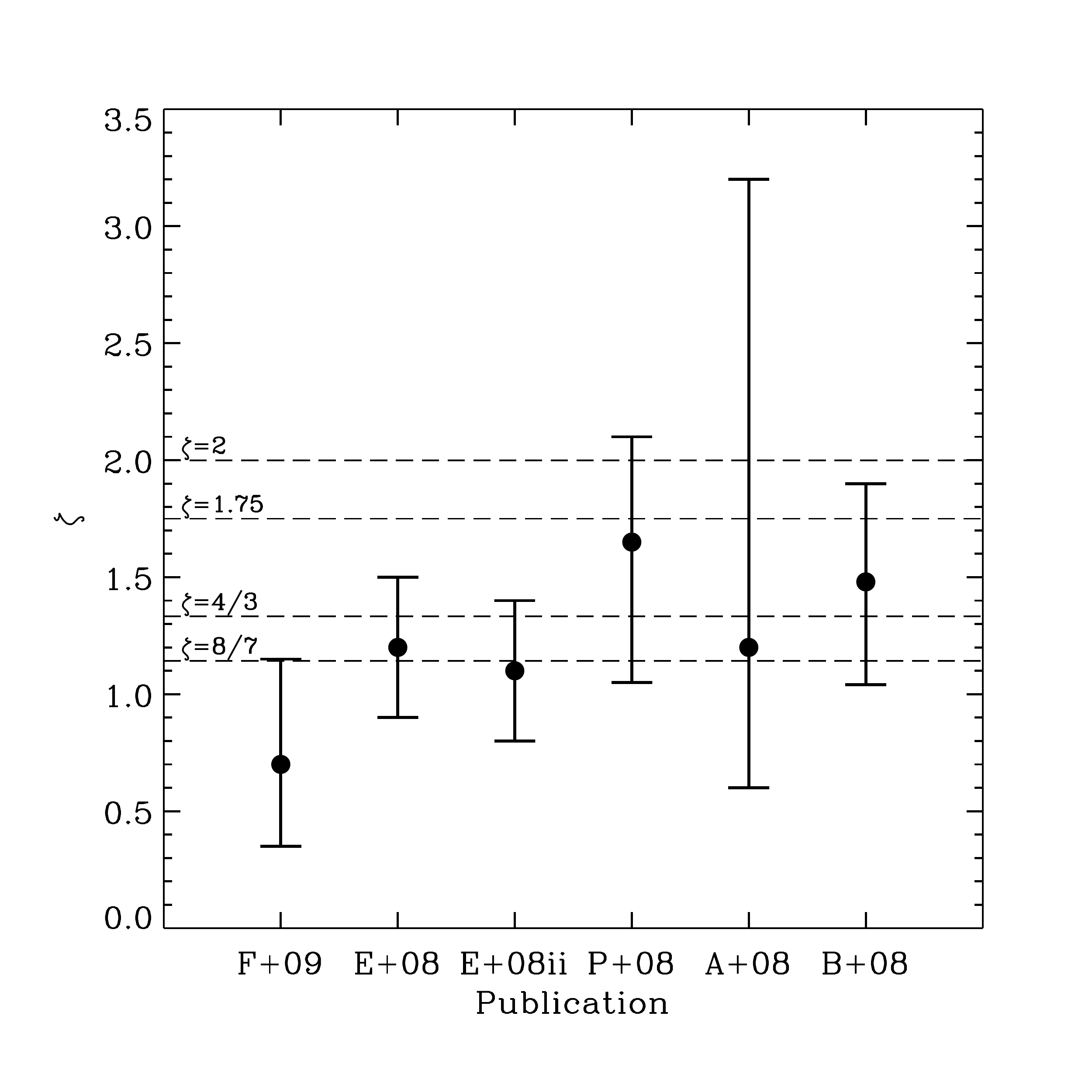}
\caption{\label{fig-accr} 
Constraints on quasar accretion mechanism in the literature (peak probability value $\pm1\sigma$). Examples shown are:
F+09 (this paper) -- \qso multi-band single-epoch imaging;
E+08~\citep{eigenbrod+08b} -- \ec~spectroscopic monitoring;
E+08ii~\citep{eigenbrod+08b} -- as E+08, but with no velocity prior;
P+08~\citep{poindexter+08} -- \he~multi-band monitoring;
A+08~\citep{anguita+08} -- \ec~multi-band monitoring;
B+08~\citep{bate+08} -- \mg~multi-band single-epoch imaging.}
\end{figure}

Results for two sources from the {\sl anomalous lens} technique are consistent. 
In Figure~\ref{fig-accr} we compare published lensing constraints on quasar accretion disc mechanisms from the literature.
It can be seen that there is a consistency amongst techniques that is apparently converging on the~\citet{agolkrolik00} model.

Data from additional new sources is required to further strengthen the results. 
Only a small number of such serendipitous targets exist, but if the current trend continues for other sources, we will be able to place strong constraints on the radial profile of the accretion disc in quasars.

\section{Acknowledgements}
DJEF acknowledges the support of a Magellan Fellowship from  Astronomy Australia  Limited. 
NFB acknowledges the support of an Australian Postgraduate Award. 
We are indebted to Joachim Wambsganss for the use of his inverse ray-shooting code. 
We gratefully acknowledge Naohisa Inada for providing us with his Magellan/MagIC images of SDSS~J0924+0219.
We acknowledge the input of an anonymous referee, who helped us clarify a number of sections of the text.

\bibliographystyle{astron}
\bibliography{anomflux}


\label{lastpage}

\end{document}